# Geometric phase from Aharonov-Bohm to Pancharatnam–Berry and beyond

Eliahu Cohen[1,2,*], Hugo Larocque[1], Frédéric Bouchard[1], Farshad Nejadsattari[1], Yuval Gefen[3], Ebrahim Karimi[1,*]

[1]Department of Physics, University of Ottawa, Ottawa, Ontario, K1N 6N5, Canada
[2]Faculty of Engineering and the Institute of Nanotechnology and Advanced Materials, Bar Ilan University, Ramat Gan 5290002, Israel
[3]Department of Condensed Matter Physics, Weizmann Institute of Science, Rehovot 76100, Israel
[*]Corresponding authors: eliahu.cohen@biu.ac.il, ekarimi@uottawa.ca

**Abstract:** Whenever a quantum system undergoes a cycle governed by a slow change of parameters, it acquires a phase factor: the geometric phase. Its most common formulations are known as the Aharonov-Bohm, Pancharatnam and Berry phases, but both prior and later manifestations exist. Though traditionally attributed to the foundations of quantum mechanics, the geometric phase has been generalized and became increasingly influential in many areas from condensed-matter physics and optics to high energy and particle physics and from fluid mechanics to gravity and cosmology. Interestingly, the geometric phase also offers unique opportunities for quantum information and computation. In this Review we first introduce the Aharonov-Bohm effect as an important realization of the geometric phase. Then we discuss in detail the broader meaning, consequences and realizations of the geometric phase emphasizing the most important mathematical methods and experimental techniques used in the study of geometric phase, in particular those related to recent works in optics and condensed-matter physics.

## 1. Introduction

A charged quantum particle is moving through space. Is it sufficient to know the local electromagnetic field in order to predict the time evolution of the particle's wavefunction? The resounding `No!' answer was given by Yakir Aharonov and David Bohm 60 years ago *(1)*. When two electronic wavepackets encircle a magnetic field which is confined to a solenoid, such that along their paths the magnetic field is zero, but the vector potential is non-zero, they will acquire a relative phase (see Fig. 1). This phase, known today as the Aharnov-Bohm phase, will affect their interference pattern when closing the loop. The phase is proportional to the enclosed magnetic flux $\Phi$ according to:

$$\varphi_{AB} = \frac{e\,\Phi}{\hbar}$$

(1)

where $e$ is the electron charge and $\hbar$ is the reduced Planck constant. The magnetic effect was anticipated in several works *(2, 3)*. However, the full electromagnetic description, as well as its significance, are is due to Aharonov and Bohm: although in classical electrodynamics potentials are regarded as mere mathematical tools, they become vital parts of the physical formalism in the quantum domain. Alternatively, the Aharonov-Bohm effect highlights a nonlocal (or global) aspect of quantum mechanics allowing electromagnetic fields to affect the charge even in regions from which they are excluded *(1,4,5,6,7,8)*. Debates regarding fundamental aspects of the Aharonov-Bohm effect remain vivid until this day *(9,10,11,12,13,14,15)*.

In the original paper *(1)* an ideal case of inaccessible, eternal and infinitely long flux line was analyzed. However, later analyses relaxing these assumptions have all reached the same conclusions *(16,17,18)*, leading Michael Berry to conclude that "The Aharonov-Bohm effect is real physics not ideal physics" *(19)*. Later on, the dual Aharonov-Casher effect was found, predicting that dipole moments diffracting around charged tubes will similarly acquire a phase *(20,21,22,23)*.

Furthermore, the Aharonov-Bohm effect has been used and demonstrated in systems such as iron whiskers *(24)*, superconducting films (thus completely excluding the magnetic field from the electron's path) *(25)*, metallic rings *(26)*, quantum dots *(27)*, carbon nanotubes *(28)*, electronic Mach-Zehnder interferometers *(29)*, topological insulators *(30,31,32)*, optical lattices *(33,34)*, and ion traps *(35)*.

Importantly, the Aharonov-Bohm phase is topological, that is it does not depend on the shape - or in general, the geometric properties - of the path, but only on its topological invariants , provided that the particle is moving in a field-free region. As deep and influential as it is, the Aharonov-Bohm phase turned out to be a special case of the broader geometric phase.

The general form of geometric phase was introduced by Michael Berry nearly 35 years ago *(36)*. He noticed that when the parameters of a quantum system are slowly cycled around a closed path, the phase of its state may not return to its original value. If the system starts in the $|n\rangle$ eigenstate of the Hamiltonian $H(\mathbf{R})$, then upon a slow change of the Hamiltonian's parameters $\mathbf{R}$, it follows from the adiabatic theorem *(37)* that at any instance of time $t$, the system will be in an eigenstate of the instantaneous Hamiltonian, that is $|n,t\rangle$. When the parameters of the Hamiltonian complete a cycle $C$, the final state will therefore return to its original value, but with an additional phase factor $\gamma[C]$, which apart from the standard dynamical contribution, depends only on the geometry of the path

$$\gamma[C] = i \int_C \langle n,t|\nabla_\mathbf{R}|n,t\rangle \cdot d\mathbf{R} \tag{2}$$

Here, $\nabla_\mathbf{R}$ is the gradient vector in the parameter space of $\mathbf{R}$. This expression is a manifestation of a holonomy in a line bundle, that is the failure to preserve geometrical data when being parallel-transported around closed loops *(38)* (see also the next section).

Having a fundamental and broad applicability *(39,40,41,42,43,44,45,46,47)*, the geometric phase was anticipated in several areas of physics *(48,49)*, most notably by Shivaramakrishnan Pancharatnam *(50)*, who showed how a geometric phase can be obtained through a sequence of (projective) measurements, specifically through measuring (and projecting) the polarization of a light beam in a cyclic manner (see also Ref. *(51)*). The geometric phase was also obtained in the context of molecular electronic degeneracies *(52)* (later generalized and further studied in *(53)*). Moreover, the geometric phase dates back to early studies concerned with conical refraction *(54, 55)*, works discussing parallel transport of the polarization along curved rays *(56, 57, 58)* and a study addressing wave propagation in stratified inhomogeneous media *(59)*.

With time, the concept of geometric phase was further generalized. The degenerate non-Abelian case was later addressed *(60)* and in this context gauge fields naturally arose. Aharonov and Jeeva Anandan rephrased the geometric phase in terms of (not necessarily adiabatic) closed loops of the quantum system itself, rather than of the Hamiltonian's parameters *(61)*. When the evolution is adiabatic, the Aharonov-Anandan phase is considered as a gauge-invariant generalization of the geometric phase. During the same year, Berry analyzed the case where the system returns to its original state only approximately, proposing an iterative technique for obtaining corrections to the geometric phase *(62)*. A further generalization was reported in a work showing that the evolution of the quantum system need be neither unitary nor cyclic *(63)*.

John Hannay proposed a classical analog known today as the Hannay angle *(64)*, which manifests itself, for instance, in the angular displacement in the position of a bob of an earth-based pendulum. For example, in the case of the Foucault pendulum, the phase shift is equal to the enclosed solid angle subtended at the Earth's center *(48,64)*. Another important manifestation of the geometric phase in solids, which will be discussed further in more detail, is known as the Zak phase *(65)*. Geometric phase also underlines the dynamics and thermodynamics of a broad spectrum of quantum systems, including those dominated by spin-orbit coupling *(66,67,68)*. As the geometric phase continues to affect all areas of physics, with an increasing number of applications, Table 1 presents only a non-exhaustive summary of all the aforementioned phases, as well as the phase originating from exchange statistics which will be discussed below. In Table 1 'topological' refers to the phase that does not depend on the geometry of the cycle. Phases that are described as non-adiabatic can appear in both adiabatic and non-adiabatic processes. In the case of the Aharonov-Bohm phase, we consider a general form for the electromagnetic potential (not just the magnetic), hence the parameter space includes both the three spatial dimensions and time.

We further discuss another extension of the geometric phase. In open (hence non-Hermitian) systems the geometric phase becomes complex *(69,70,71,72,73,74,75,76)*; its real part is modified as compared with the corresponding closed system, and its imaginary part (which has a geometric character too) gives rise to 'geometric dephasing' *(77,80,81)*. This includes coupling to a dissipative *(77)* or noisy *(78,79)* environment.

Notwithstanding the energy spectrum of the system and environment is continuous, one may still define the geometric phase and compute its variation, averaging over a distribution of closed trajectories. Following a closed trajectory adiabatically (over time $t$) $n$ times ($n$ being the winding number), the initial wavefunction $|\psi\rangle_{\text{initial}}$ is multiplied by a factor, given schematically by

$$|\Psi_{final}\rangle = e^{iET|n|-\alpha T|n|}e^{i\theta^0_{Berry}T^0n+i\delta\theta^{Re}_{Berry}T^0n-\delta\theta^{Im}_{Berry}T^0n}e^{(\ldots)T^{-1}|n|}|\Psi_{initial}\rangle \qquad (3)$$

The terms in the exponential factors represent an expansion in powers of $1/T$. Here, the first factor, proportional to $T^1$, corresponds to the dynamical phase and dynamical dephasing. The second term, $\sim T^0$, is geometric in nature (hence the arguments are proportional to $n$). The environment-induced correction to the geometric phase is complex: this leads to a correction of the magnitude of the geometric phase, and, at the same time, to real terms in the exponential, $e^{-\delta\theta^{Im}_{Berry}T^0 n}$, dubbed 'geometric dephasing' *(80)*. The latter may have either $\pm$ signs, depending on the direction of the winding, thus enhancing or reducing the standard dynamical dephasing. This effect has been observed experimentally *(81)*. Terms with powers of $1/T$ (with the winding factor alternating between $|n|$ and $n$, correspond to non-adiabatic corrections. The $e^{(\ldots)T^{-1}|n|}$ term is a non-universal constant.

Interestingly, the geometric phase can have a topological character in non-Hermitian systems *(82,83,84)*.

## 2. Mathematical formalism

The distinguishing feature of geometric phases is that, unlike dynamical phases, they are not associated with the action exerted by the potentials throughout a displacement in space. In other words, these phases are not attributed to the forces applied onto the quantum system. Instead, they are associated with the connection of space itself. The Aharonov-Bohm phase is a special case of the geometric phase as it consists of the Dirac phase, $(e/\hbar)\int_\gamma \mathbf{A}\cdot d\mathbf{r}$ – where $e$ is the electron charge and $\hbar$ is the reduced Planck constant – acquired by a quantum system upon being displaced along a trajectory $\gamma$ located in a manifold where there is a vanishing magnetic field attributed to a vector potential **A** *(1)*. This phase, however, can be described within a much more general quantum mechanical framework in which geometric phases arise purely from the adiabatic evolution of a state $|n(\mathbf{R})\rangle$ upon the modification of the parameters **R**. At this point, the evolution of the state becomes associated with the Berry connection $\mathcal{A}=i\langle n(\mathbf{R})|\nabla_\mathbf{R}|n(\mathbf{R})\rangle$ giving rise to a geometric phase of $\int_\gamma \mathcal{A}\cdot d\mathbf{R}$, that, as emphasized for example in Refs. *(52,53,85)*, will even act back on the parameters driving it with a Lorentz-like force depending on the speed at which the parameters are changed. This connection consists of a generalization of the vector potential in the Aharonov-Bohm phase. In fact, a direct calculation of this quantity ties it to the electromagnetic vector potential *(36)*, where **R** in this case is the origin of the displaced quantum state, thereby confirming that the Aharonov-Bohm phase is indeed geometric *(86)*. However, in some cases the interplay between the Aharonov-Bohm phase and the Berry phase turns out to be more intricate *(87)*.

Geometric phases are deeply connected to the topological structure formed by a quantum system's Hilbert space, $\mathcal{H}$, and the space of its adiabatically varied parameters **R**, *M* Ref. *(38)*. This structure is formally known as a vector bundle which consists of the manifold *M* embedded within the space $M\times\mathcal{H}$. The term bundle refers to the collection of the different Hilbert spaces defined for specific values of **R**, each of them connected to the parameter space *M* by means of the wavefunctions $\psi(\mathbf{R})$. The wavefunctions mathematically comprise the sections of the vector bundle and differentiating them provides a means for gauging their evolution within $M\times\mathcal{H}$. However, this differentiation is not merely as simple as taking the derivative of the wavefunction, given that they are not similarly defined from one Hilbert space to the other. Instead, one must use what is known as a connection, *D* (see ref. *(88)*), which is often associated to vector potentials, $\mathcal{A}$. In fact, all vector potentials defined over a vector bundle must differ by a connection much like how all vector potentials attributed to an electromagnetic field are related to one another by a gauge transformation. For this reason, the terms connections and vector potentials are often used interchangeably by many physicists who refer to $\mathcal{A}$ as the bundle's connection (we will use the term bundle's connection in this review).

Connections formally allow the differentiation of wavefunctions over **R** by providing a unique way of dragging them from one Hilbert space in $M\times\mathcal{H}$ to another. This process is known as parallel transport and can occur provided that a smooth path between both spaces, $\gamma$, is defined. The smoothness of $\gamma$ is crucial to the transport of $\psi(\mathbf{R})$ and is quantum mechanically enabled by the adiabatic evolution of wavefunctions *(36,38)*. Due to the topological structure of $M\times\mathcal{H}$, parallel transport along a closed path will map a state $\psi(\mathbf{R})$ to $H(\gamma,D)\psi(\mathbf{R})$, where $H(\gamma,D)$ is a linear map known as the holonomy of the path (Fig. 2). In fact, the holonomy precisely corresponds to the geometric phase term added to $\psi(\mathbf{R})$ upon experiencing a cyclic variation of **R**.

Another important property of the vector bundles is their curvature, which is related to many of the aforementioned quantities. Namely, it can be shown that the curvature corresponds to the holonomy of an infinitesimally small loop. Moreover, the way in which the curvature is related to connections is highly reminiscent of how magnetic fields are equal to the curl of the electromagnetic vector potential. And much like the magnetic field itself, it is uniquely defined in spite of there being several formulations of the bundle's connection. Finally, the curvature integral over a surface bounded by its holonomy's closed loop precisely yields the geometric phase. This integral, however, also holds a deeper topological meaning as it essentially yields an integer *(38)*. This integer determines the Chern class *(89)* of the bundle's connection and provides details pertaining to the topological structure of the bundle such as the strength of singularities that it could be containing *(90,91,92,93,94,95,96,97,98,99)*.

### 3. Geometric phase in optics

There are two main types of geometric phase in optics *(48,100)*. The first, 'spin-redirection' geometric phase, commonly refers to light with fixed state of polarization changing direction. The second, the Pancharatnam-Berry phase, usually applies to light propagating through anisotropic medium in a fixed direction with a slowly changing state of polarization. Below we shall focus on the latter, but the former - first studied in *(56,57,58)* and tested in *(101,102,103)* -- fundamentally reflects intrinsic spin properties of relativistic wave equations *(104,105,106)* and has been as a source of major interest.

*Pancharatnam-Berry geometric phase*

One of the first observations of the geometric phase was in the field of optics by Pancharatnam *(50)*. He noticed a phase shift arising when the polarization of a photonic beam was varied in a cyclic manner. The Poincaré sphere representation of polarization states was a key element to help Pancharatnam recognize the geometric nature of the arising phase shift. This geometrical visualization, indeed, lies in the fact that the 2-dimensional polarization vector space can be mapped onto a surface of a sphere, known as Poincaré or Bloch sphere. The Poincaré sphere representation allows one to visualize pure polarization states on a unit sphere (Fig. 3). Though this mapping is somewhat arbitrary, points of circular polarization are conventionally located at the north and south poles. Thus, states of linear polarization are located along the equator, and other elliptical polarization states are located elsewhere on the sphere. Moreover, the Poincaré sphere representation may also account for partially polarized beams by locating them inside the unit sphere, with the extreme case of completely unpolarized light located at the centre of the sphere.

Another useful representation of polarization, also employed in Pancharatnam's work, is known as the Jones vector formalism *(107)*. Here, polarization states are represented by a normalized bi-dimensional complex vector $\left(E_x\, e^{i\phi_x}\ E_y\, e^{i\phi_y}\right)^T$, where $x$ and $y$ are transverse Cartesian coordinates, and $T$ represents transposition. $E_x$ and $\phi_x$ represent, respectively, the amplitude and phase of the $x$-component of the electric field, and likewise for the $y$-component. Thus, the Jones vectors of horizontal, vertical, diagonal, anti-diagonal, left-, and right-circular polarizations are respectively given by $|H\rangle = (1\ 0)^T$, $|V\rangle = (0\ 1)^T$, $|D\rangle = (1\ 1)^T/\sqrt{2}$, $|A\rangle = (1\ -1)^T/\sqrt{2}$, $|L\rangle = (1\ i)^T/\sqrt{2}$, and $|R\rangle = (1\ -i)^T/\sqrt{2}$. The link between the Jones vectors and the Poincaré sphere representation is achieved through the reduced Stokes parameters, $s_1$, $s_2$, and $s_3$. In the Poincaré sphere representation, the reduced Stokes parameters correspond to the $x$, $y$, and $z$ Cartesian coordinates. In the Jones vector formalism, the reduced Stokes parameters are given by, $s_1 = P_H - P_V$, $s_2 = P_D - P_A$, and $s_3 = P_L - P_R$, where $P_H$, $P_V$, $P_D$, $P_A$, $P_L$, and $P_R$ correspond to the absolute square of the overlap with the horizontal, vertical, diagonal, anti-diagonal, left, and right-circular polarization states, respectively. Thus, the reduced Stokes parameters offer the link between the mathematical formalism and the geometric nature of the polarization of light.

Pancharatnam introduced the notion of phase difference between two polarization states, $|\psi_1\rangle$ and $|\psi_2\rangle$, as the phase of their scalar product, that is $\arg[\langle\psi_1|\psi_2\rangle]$. He then noted the non-transitivity of the phase difference among three polarization states. For instance, if the polarization states $|\psi_A\rangle$ and $|\psi_B\rangle$ are in phase, and $|\psi_B\rangle$ and $|\psi_C\rangle$ are also in phase, it is not necessary for $|\psi_C\rangle$ and $|\psi_A\rangle$ to also be in phase; the following set of polarization states: $|\psi_A\rangle = |L\rangle$, $|\psi_B\rangle = |H\rangle$, and $|\psi_C\rangle = |A\rangle$ describes the above situation. In fact, if the state $|\psi_C\rangle$ is indeed in phase with another polarization state $|\psi_{\tilde{A}}\rangle$ located at the same point as $|\psi_A\rangle$ on the Poincaré sphere, the phase difference between $|\psi_A\rangle$ and $|\psi_{\tilde{A}}\rangle$ will be given by $\arg[\langle\psi_{\tilde{A}}|\psi_A\rangle] = \Omega_{ABC}/2$, where $\Omega_{ABC}$ is the solid angle of the geodesic triangle spanned by the points on the Poincaré sphere corresponding to the states $|\psi_A\rangle$, $|\psi_B\rangle$, and $|\psi_C\rangle$. Berry subsequently showed the equivalence between Pancharatnam's geometric phase in polarization and the broader phenomenon of the acquired phase in the adiabatic evolution of a state in quantum mechanics *(51)*.

One of the most important instances where we observe geometric phases in polarization is the case of birefringent materials *(108,109,110)*. In particular, a half-wave plate (HWP), is a piece of birefringent material that induces a relative phase of half a wavelength between the two polarization components aligned with the slow and fast axes of the plate. When a linearly polarized beam is made to pass through a HWP, the exiting beam is given by a linearly polarized beam rotated by an angle $2\theta$ from the polarization orientation of the input beam's polarization, where $\theta$ is the angle of rotation of the HWP with respect to its fast axis (optical axis). In the case of an input circularly polarized beam, the HWP has the effect of flipping the handedness of the circular polarization state, that is from left to right-circularly polarization and from right to left-circularly polarization. This flip of handedness is independent of the orientation of the HWP. However, the outgoing beam acquires an additional global phase that is given by $2\theta$, where $\theta$ is once again the angle of rotation of the HWP. This extra phase factor is exactly the geometric phase observed by Pancharatnam when the polarization state evolves from left- to right-circular polarization, and vice-versa (Fig. 3-c,d).

*Optical phase elements*

The geometric phase obtained from passing a beam of circularly polarized light through a HWP is very useful in terms of wavefront shaping *(111,112,113,114)*. One can imagine segmenting a HWP into several parts. By carefully adjusting the orientation of each part of the HWP, one can precisely shape the wavefront of a beam passing through such a device. For instance, by varying the orientation of the HWP in terms of the transverse position, for example, the $x$ coordinate, the resulting linear phase shift will tilt the wavefront of the incoming beam and redirect the wave-vector of the beam towards the left (Fig. 3-a). However, the geometric phase offers more than just shaping the wavefront of an incoming beam. It simultaneously shapes the wavefront of a beam in a polarization-dependent manner. In the case described above, the segmented HWP with a rotation angle linearly depending on the transverse coordinate $x$ will redirect a left and a right-circularly polarized beam toward the right and left, respectively, see Fig. 3-b. Moreover, this whole process is fully coherent and thus valid for any input beam comprised of a superposition of left- and right-circularly polarized light. More specifically, by manipulating the local orientation of the optical axis, $\theta(x, y)$, of a birefringent medium with an optical retardation of half a wavelength, it is possible to respectively imprint a phase pattern of $\exp[2i\theta(x,y)]$ and $\exp[-2i\theta(x,y)]$ for an input beam with left and right circular polarization. For instance, a polarization-dependent focusing lens has been designed using geometric phases *(115)*. This type of wavefront shaping has also been investigated in several types of devices *(116,117,118,119)*.

It is worth noting that other manifestations of the geometric phase exist in optics when other photonic degrees of freedom are considered. The modal structure of light provides an unbounded vector space, and thus can be mapped onto the surface of a hypersphere embedded in $\mathbb{R}^d$, where d is the dimension of the vector space or onto a surface of higher-order Poincaré spheres *(120,121,122,123,124,125)*.

A particular class of wavefront shaping that has attracted significant attention in the last decade is the generation of beams carrying orbital angular momentum (OAM) *(126)*. It has been recognized that vortex beams with a phase term $\exp(i\ell\phi)$, where $\ell$ is an integer number denoting the OAM value and $\phi$ is the azimuthal coordinate, carry a quantized OAM given by $\ell\hbar$ per photon *(127)*. OAM carrying beams have since been generated experimentally using refractive elements, that is spiral phase plates *(128)*, and pitch-fork holograms *(129,130)*. In 2006, a device called the $q$-plate was introduced and was capable of generating OAM in the visible domain using a patterned liquid crystal cell *(131)*. The $q$-plate works precisely under the principle of geometric phases. It consists of a birefringent medium with a tunable optical retardation which is typically set to half a wavelength. The orientation of the local optical axis of the liquid crystal device is given by $\theta(\phi) = 2q\phi$, where $q$ is called the topological charge of the $q$-plate. In order to generate a beam carrying an OAM value of $\ell$, a topological charge of $q = \ell/2$ is imprinted. The generation of OAM-carrying beams using the $q$-plate has also been recognized as being the result of spin-orbit coupling in an inhomogeneous medium *(132, 133)*. In addition to carrying OAM, light may also carry spin angular momentum (SAM). For instance, a left- and right-circularly polarized beam respectively carry a SAM value of $+\hbar$ and $-\hbar$ per photon. Let us consider the specific case of $q = 1$, where the $q$-plate possesses cylindrical symmetry. For an input left-circularly polarized beam with zero OAM, the total angular momentum of the beam is given by the sum of its SAM an OAM, hence $+\hbar$ per photon. After passing through a $q = 1$-plate, the polarization is flipped to right-handed circular polarization, and the OAM is increased to $\ell = 2q = 2$. Thus, the total angular momentum is now given by $-\hbar + 2\hbar = \hbar$. As we can see, the total angular momentum is conserved, and the process of wavefront shaping via geometric phases can be seen as a form of spin-orbit coupling *(131,134)*. Since then, spin-orbit coupling devices manipulating geometric phases have been demonstrated in a wide range of applications *(135)*. Moreover, the spin-orbit-coupling resulting from the $q$-plate (or other structured plates) has allowed the generation of exotic polarization structures, for example, vector vortex beams *(136,137,138)*, Poincaré beams *(139)*, and 3-dimensional polarization structures *(140)*. Meta-surfaces with a birefringence resulting from plasmonic resonances have been experimentally demonstrated as a means to generate OAM at the integrated level using devices with thicknesses that can be as small as $1/30$ of a wavelength *(119,141,142)*, see Fig. 4. Spin-orbit coupling due to geometric phase can also occur at the interface of two materials. Due to the transverse nature of light and the need to satisfy boundary conditions at an interface between two materials, for example, air and glass, slight variation in a beam's propagation direction can be observed depending on the SAM of the incoming beam. This effect can be seen as a result of spin coupling to the transverse momentum of a light field.

Further manifestations of geometric phases in spin-orbit coupling has been observed in evanescent waves upon total internal reflection. These effects have been investigated theoretically and observed experimentally under various circumstances *(143,144,145,146)*.

Using this optical spin-orbit coupling resulting from a geometric phase, several experiments have demonstrated their uses in quantum information and quantum simulation tasks. A striking example is the linear quantum walk performed using OAM states of photons as the walker space, and the polarization of the photons as the coin in the quantum walk *(147,148,149,150)*, see Fig. 5. Taking advantage of the inherent coherence of the geometric phases, a coin prepared in a balanced superposition of left- and right-circular polarization, that is linear polarization, will cause a walker to move up and down in its OAM space, coherently, when passing through a $q$-plate. This way, one may build an in-line quantum walk setup by cascading several $q$-plates and using the appropriate polarization optics. The number of $q$-plates corresponds to the number of steps in the quantum walk and scales linearly with the number of elements. Quantum walks have been shown to be promising tools for quantum simulations and quantum computations *(151,152)*. Thus, one can appreciate the breadth and impact that geometric phases currently may have in many different areas of optics.

Several of the above spin-orbit effects, which are geometric in nature, have been translated to the physics of electron beams. Most notably, deflections in an electron's motion caused by electromagnetic fields can be seen as a source of adiabatic variations in the electron's longitudinal momentum vector, which turn into adiabatic variations in its spin, thereby leading to a geometric connection that influences its dynamics *(153)*. For the cases of specially structured electromagnetic fields, these deflections can lead to the gain of an azimuthally-dependent geometric phase *(154)* akin to that acquired by a photon propagating through a $q$-plate.

## 4. Role in condensed-matter physics

As far as a single particle picture is concerned, the Berry connection and curvature may appear over a 7D parameter space including the 4D spacetime and 3D momentum space. The corresponding components of the Berry curvature over the 7D parameter space contribute to the general semi-classical wave-packet equations of motion *(47,155,156)*. Below we discuss a full-fledged quantum picture of some paradigmatic manifestations of geometric phase in condensed matter. Some important themes (for example, the relation between electrons' spin-orbit interaction and geometric phase *(66)*) are omitted here.

*Electronic Bloch states*

For electrons in simple crystalline systems (with no impurities and defects) such as periodic lattices, regardless of the specific form of the potential and only due to the periodicity of the Hamiltonian (translational symmetry), Bloch's theorem is applicable. Based on this theorem, the energy eigenstates of the underlying periodic Hamiltonian known as Bloch states, are given by $|\psi_{n,\boldsymbol{k}}(\boldsymbol{r})\rangle = e^{i\boldsymbol{k}\cdot\boldsymbol{r}}|u_{n,\boldsymbol{k}}(\boldsymbol{r})\rangle$. Here, $n$ is the energy band index, $\boldsymbol{k}$ is the electronic quasi-momentum (in units of $\hbar$), and $u(\boldsymbol{r})$ is a periodic function with the periodicity of the underlying Bravais lattice vector, $\boldsymbol{R}$, $u(\boldsymbol{r} + \boldsymbol{R}) = u(\boldsymbol{r})$. Consider a single, isolated, band. Following $\boldsymbol{k} = \boldsymbol{k}(t)$ throughout the Brillouin zone, the Bloch state of an electron traverses a closed path in parameter space (the reciprocal quasi-momentum space). This is where the geometric phase emerges. Although within a unit cell of the crystal, quasi-momenta differing by a reciprocal lattice vector, $\Delta \boldsymbol{k} = \boldsymbol{G}$, are considered equivalent, Bloch states are characterized by an additional quantum number, the band number $n$, leading to $n$-dependent non-dynamical geometric phases $\gamma_n$:

$$|\psi_{n,\boldsymbol{k}(t)}(\boldsymbol{r},t)\rangle = e^{i\gamma_n} e^{-\frac{i}{\hbar}\int_{t_0}^{t} \mathcal{E}_n(\boldsymbol{k}(t'))dt'} |\psi_{n,\boldsymbol{k}(t)}(\boldsymbol{r},t_0)\rangle \tag{4}$$

where $e^{-\frac{i}{\hbar}\int_{t_0}^{t} \mathcal{E}_n(\boldsymbol{k}(t'))dt'}$ is the dynamical phase associated with the evolution of the eigenstate $|\psi_{n,\boldsymbol{k}(t_0)}\rangle$, and $\mathcal{E}_n(\boldsymbol{k})$ is the corresponding energy eigenvalue. The adiabatic evolution of the original Bloch state $|\psi_{n,\boldsymbol{k}}(\boldsymbol{r},t_0)\rangle$ through a closed loop in $k$-space (the first Brillouin zone), replaces now the aforementioned abstract set of parameters $\boldsymbol{R}$. This results in the geometric phase, $\gamma = \oint \boldsymbol{\mathcal{A}}(\boldsymbol{k}) \cdot d\boldsymbol{k}$, where $\boldsymbol{\mathcal{A}}(\boldsymbol{k})$ is the Berry connection, which for a given energy band is defined as: $\boldsymbol{\mathcal{A}}(\boldsymbol{k}) = -i\langle u_{\boldsymbol{k}}|\nabla_{\boldsymbol{k}}|u_{\boldsymbol{k}}\rangle$. It must be noted that the Berry connection is a gauge-dependent parameter and thus does not correspond to an observable, however the geometric phase defined as an integral over a (closed) loop is gauge invariant. It thus corresponds to a physical observable. Another important quantity here is the curl of the Berry connection, known as the Berry curvature: $\boldsymbol{\Omega}(\boldsymbol{k}) = \nabla_{\boldsymbol{k}} \times \boldsymbol{\mathcal{A}}(\boldsymbol{k})$. Physically, the Berry curvature (which, similarly to $\boldsymbol{\mathcal{A}}(\boldsymbol{k})$, is gauge-dependent), is a measure of the local rotation of the electronic wavepacket as the latter traverses the Brillouin zone. By means of the Stokes' theorem one can write the geometric phase as an integral over a manifold of the Berry curvature. If the manifold is closed (such as a two-dimensional torus in reciprocal space), then the result is topological and quantized by a $2\pi$ multiple of the so-called Chern numbers. In 1989 Joshua Zak has first introduced the concept of geometric phase for Bloch electrons in one-dimensional periodic lattices *(65)* (Fig. 6-a). Zak has shown that in the case of an asymmetric lattice, Bloch electrons can be accompanied by a non-trivial geometric phase, known nowadays as the Zak phase. In the presence of inversion symmetry, the corresponding Zak phase can assume only the trivial values of 0 or any integral multiple of $2\pi$. In a three-dimensional lattice, when inversion symmetry is present, the closed loop along a given symmetry axis in parameter space will convert into a three-dimensional torus representing the Brillouin zone and the corresponding geometric phase will take well-defined and quantized values that correspond to the Wyckoff positions *(157)* (which are a set of spatial coordinates related to the point group and space group symmetries of the underlying crystal lattice). Likewise, in a two-dimensional lattice, the Brillouin zone is regarded as a closed two-dimensional torus, and the corresponding geometric phase will thus be an integral of the Berry curvature over a two-dimensional manifold. Zak's main conclusion was that the geometric phase can be used to characterize all the energy bands in solids. In the case of multiple bands, one can define the Zak phase as the sum over all phases $\gamma_n$ related to each of the individual and occupied energy bands:

$$\gamma = \sum_n \gamma_n \tag{5}$$

*Quantum Hall Effect*

The quantum Hall effect is arguably the paradigmatic topological phase in condensed matter physics. There are two complementary ways to approach the topological facets of this physics: analyzing bulk properties (this was formulated and studied primarily through the well-known Thouless, Kohomoto, Nightingale and den Nijs (TKNN) paper *(158)*, and focusing on edge physics. For example, one may derive two equivalent expressions for the transverse Hall conductance, focusing on bulk physics and on edge physics respectively.

It is known that electrons confined to a plane in low temperatures and in the presence of strong magnetic fields show (quantized) plateaus in the Hall conductance. The TKNN formalism asserts that the topological Hall conductance is closely related to the Berry curvature of the Bloch states. One considers a magnetic field perpendicular to a plane that confines a two-dimensional electron gas. In addition, a relatively weak in-plane electric field is applied along the *y*-direction. One may then define the longitudinal conductance, $\sigma_{xx} = \frac{j_x}{E_x}$, as well as the transverse Hall conductance, $\sigma_{xy} = \frac{j_x}{E_y}$, where $j_{x(y)}$ and $E_{x(y)}$ are the current densities and the electric fields along the respective directions. Starting from the current density $\boldsymbol{j} \equiv -\frac{1}{V}\sum_{n,k} e\boldsymbol{v}_n(\boldsymbol{k})$, one can derive the Hall conductance. Here, $\boldsymbol{v}_n(\boldsymbol{k}) = \langle \psi_{n,k} | \frac{1}{\hbar}\frac{\partial H}{\partial \boldsymbol{k}} | \psi_{n,k} \rangle$ is the velocity of the particles in the *n*th band, where $H$ is the k-dependent Hamiltonian, and $|\psi_{n,k}\rangle$ are the Bloch states. In the absence of an external electric field, $\boldsymbol{v}_n(\boldsymbol{k}) = \frac{1}{\hbar}\frac{\partial \mathcal{E}_n(\boldsymbol{k})}{\partial \boldsymbol{k}}$ is the group velocity at $(n, \boldsymbol{k})$. Employing modified semi-classical equations of motion for Bloch electrons in the presence of an external electric field, $\boldsymbol{E}$, that is. $\boldsymbol{v} = \dot{\boldsymbol{r}} = \frac{1}{\hbar}\frac{\partial \mathcal{E}(\boldsymbol{k})}{\partial \boldsymbol{k}} - \dot{\boldsymbol{k}} \times \boldsymbol{\Omega}(\boldsymbol{k}), \dot{\boldsymbol{k}} = \frac{e}{\hbar}\boldsymbol{E}$, the current density (omitting the drift group velocity term for the two-dimensional electron gas) reads:

$$\boldsymbol{j} = -\frac{1}{V}\sum_{n,k} e\boldsymbol{v}_n(\boldsymbol{k}) = \frac{e^2}{\hbar}\int \frac{d^2\boldsymbol{k}}{(2\pi)^2}\sum_n \boldsymbol{\Omega}_n(\boldsymbol{k}) \times \boldsymbol{E}$$

(6)

Thus, one can obtain the components of the electric conductance tensor $\boldsymbol{\sigma}$ from $\boldsymbol{j} = \boldsymbol{\sigma}\boldsymbol{E}$. In particular, the component of the electric current density perpendicular to the Hall field in this model, $j_x$, leads through the following relation $j_x = \frac{e^2}{\hbar}\int \frac{dk}{2\pi}\sum_n \Omega_z^n E_y$ to the transverse Hall conductance $\sigma_{xy}$:

$$\sigma_{xy} = \frac{e^2}{\hbar}\sum_n \int \frac{dk}{2\pi}\Omega_z^n$$

(7)

where, $\Omega_z^n$ represents the *z* component of berry curvature, $\boldsymbol{\Omega}_n(\boldsymbol{k})$.

The Chern number, which is (up to a factor of $1/2\pi$) the integral of the Berry curvature over a two-dimensional Brillouin zone, is an integer. The Hall conductance is an integral multiple of $e^2/h$, and is a topological property of the system. This invariant quantity can be defined only if the bands are completely full.

A similar analysis can be applied to the motion of electrons at the edge of a 2D or 3D structure, see Fig. 6b, thereby revealing the presence of an 'anomalous Hall effect' related to the broken time reversal symmetry in these phases of matter *(159,160,161)*.

*Electric polarization*

A unifying framework for understanding the macroscopic polarization of crystalline dielectrics was not conceivable before the incorporation of geometric phase. The bulk Bloch electron density, $\rho(\mathbf{r}) = |\psi_{n,\mathbf{k}}(\mathbf{r})|^2$, is not sufficient for determining the electric polarization vector of the material, **P**. This ill-defined electric polarization in infinite lattices and lattices with periodic boundary conditions posed a major problem in the theory of solids. The modern theory of electric polarization relies on including the geometric phase effects *(162,163)*. The underlying freedom with which one can define the boundaries of the unit cell in a given crystal lattice with respect to the atomic basis, that is its charge distribution, is the origin of the problem (Fig. -c). However, the change in polarization, $\Delta \mathbf{P}$ is a well-defined quantity, independent of one's definition of the unit cell *(162)*. The change in polarization is directly related to the geometric phase of the electronic states in periodic structures. This is understood within a single-particle picture, as well as within the Kohn-Sham schemes incorporating orbitals generated in density functional theory *(164,165)*. Consider the variation of the polarization upon adiabatic evolution of the displacement in the ions' positions in a periodic lattice. It can be represented by: $\Delta \boldsymbol{P} = \int_0^1 \frac{d\boldsymbol{P}}{d\lambda} d\lambda$, where $\lambda$ is a perturbation parameter indicating the gradual displacement of the ions under an adiabatic evolution. The polarization vector is defined as: $= \frac{e}{V} \sum_i \langle \psi_i | r | \psi_i \rangle$, where the sum runs over all filled Bloch states, and $V$ is the volume of the solid. It can be shown that the dependence of the polarization on the perturbation parameter is given by:

$$\frac{dP_j}{d\lambda} = \frac{e}{V} \sum_{n,\boldsymbol{k}} \Omega^n_{k_j,\lambda}(\boldsymbol{k})$$

(8)

here $j = x, y, z$ represent the three Cartesian coordinates, and the $\Omega_j$ parameters are the components of the Berry curvature defined in this case as *(163)*:

$$\Omega^n_{k_j,\lambda} = i \left( \left\langle \frac{\partial u}{\partial k_j} \middle| \frac{\partial u}{\partial \lambda} \right\rangle - c.c. \right)$$

(9)

Here, $u$ stands for a cell-periodic function in the Bloch states, as defined above. In the case of an electron propagating in a one-dimensional lattice with a lattice constant of unity, the expression for the polarization vector simplifies to:

$$\Delta P = \frac{e}{2\pi} \int_0^1 d\lambda \sum_n \int_0^{2\pi} dk \, \Omega^n_{k,\lambda}$$

Now using Stokes' theorem one can convert the double-integral over the area in the $k$-$\lambda$ space to a one-dimensional integral over the boundaries, thus relating the change in polarization to a Zak phase:

$$\Delta P = \frac{e}{2\pi} \sum_n \gamma_n$$

(11)

Here the summation runs over all the occupied states. An important implication of this result is known in cases where the initial volume of the unit cell is equal to the volume after the perturbation was applied. In these situations, the Zak phase can only be an integer multiple of $2\pi$ as mentioned earlier. An alternative and equivalent approach to the problem of polarization employs Wannier functions *(166)*. Generalizing to higher dimensional lattices such as two and three-dimensional systems, the components of $\Delta P$ are directly proportional to the corresponding components of the Berry curvature and are given by:

$$\Delta P_j = \frac{ef}{(2\pi)^N} \sum_n \int_0^1 d\lambda \int d^N k \ \Omega^n_{k_j,\lambda}$$

(12)

where *f* is the occupation number of states (in the valence band) and *N* is the dimensionality of space. The summation runs over all occupied bands, and the integral over *k*-space spans the corresponding (higher dimensional) Brillouin zone.

*Exchange statistics*

There is a deep connection between exchange statistics of spins and the geometric phase. When fully incorporating identicalness into the Hilbert space, the Pauli sign $(-1)^{2S}$ was derived, where *S* is the spin quantum number, as a geometric phase factor of topological origin *(167,168)*. The exchange of two identical particles, *i* and *j*, belonging to a many-body state, $\Psi$, is represented by the operator $P_{i,j}$. For Abelian particles, the outcome of the exchange operation can be a phase $\theta_{ij}$, namely:

$$P_{i,j}\Psi(x_1, x_2, \ldots, x_i, \ldots x_j \ldots) = \Psi(x_1, x_2, \ldots, x_i, \ldots x_j \ldots) = e^{i\theta_{ij}}\Psi(x_1, x_2, \ldots, x_i, \ldots x_j \ldots)$$

(13)

Doubly exchanging the particles amounts to rotating one particle around the other. In three spatial dimensions, the trajectory representing this rotation can be continuously shrunk to a point, hence this braiding degenerates to the unit operator, $P^2_{i,j} = 1$. This implies that $\theta_{ij}$ must be either equal to 0 or $\pi$ (up to integer multiples of $2\pi$), defining bosons and fermions respectively. This turns out not to be the case in 2D, where the trajectory of particle $i$ circumventing particle $j$ cannot be brought continuously to a point. One may then have a generalization of these two classes of particles *(169)*: particles which are not necessarily bosons or fermions. Indeed, they may obey any statistics and thus are called anyons *(170)*. We note that the geometric topological phase accumulated over the closed trajectory (of $i$ circumventing $j$) is $\mp 2\theta_{ij}$, where the sign is dictated by the direction of the winding (clockwise or counter-clockwise respectively). This is in contrast to the standard cases of bosons and fermions, where the sign of the winding is immaterial (that is, the value of $2\theta_{ij}$ does not depend on the direction of the winding). The fact that for 2D anyons $P_{i,j}$ depends on the 'history' of the exchange (that is, on details of the trajectory of the particle exchange) (, Fig. 7), implies that a simple-minded second quantized form of anyons is unattainable. One can nevertheless define anyonic Green functions by attaching a magnetic fluxon to the charged anyons (see for example, *(171)*), by implementing an exchange convention (for example, always rotate clockwise) *(172)*, or by addressing an anyonic creation/annihilation product which is insensitive to the sign of $\theta_{ij}$ *(173)*.

Obvious realizations of anyons are quasi-particles of fractional quantum Hall phases *(174)*. These particles possess fractional charge which may, under certain conditions, lead to fractional Aharonov-Bohm oscillations *(175)*, and, at the same time fractional (anyonic) exchange statistics. Theoretical works proposed various interference platforms as protocols to detect fractional statistics *(176)*, with the leading candidate being electronic Mach-Zehnder interferometers *(173,177)*. There are also proposals for the detection of non-Abelian exchange statistics *(177,178)*, where following a closed braiding trajectory the system's many-body state is modified (that is, not only multiplied by a phase, but rather rotated in a degenerate Hilbert subspace). Notwithstanding extensive efforts, clear observation of a fractional statistical phase remains an open challenge.

## 5. Conclusion

The geometric phase is a deep yet widely-applicable concept affecting many areas of physics chemistry, mathematics, engineering and computer science. We have outlined a few well-known manifestations of the geometric phase (summarized in Table 1) and briefly discussed its general mathematical formalism, before focusing on its manifestations within optics and condensed-matter physics. Alongside with the latter fields of research, the geometric phase has also influenced high energy and particle physics *(179,180,181)*, gravity and cosmology *(182,183,184)*, fluid mechanics *(185,186)*, chemical physics *(41,187,188,189)* and many other areas of active research. In optics, geometric phases give rise to unprecedented forms of optical dynamics. They also provide a new platform for optical wavefront shaping, thereby leading to various constructs for quantum simulators. In solid-state physics, geometric phases most notably provide methods for examining the band structure of materials, thereby leading to the observation of topological phenomena in condensed matter. Geometric phase phase can be defined and has been measured (for example, *(81)*) in open dissipative systems, leading to the concept of geometric dephasing. The broad implications of the latter, especially for quantum processing setups, are yet to be understood and so are the implications of a recent connection with random matrix theory *(190)* for disordered systems. Finally, we wish to mention the promising role of geometric phase and in particular non-Abelian geometric phase in quantum information and computation *(191,192,193,194,195,196,197,198,199)*. Implementations of quantum gates based on the geometric phase (in its degenerate non-Abelian form *(60)*), as first suggested by *(191)*, may lead to stable, large scale and fault-tolerant quantum computation. The main reason is that all errors usually arising from the dynamical phase will be naturally eliminated by this method. In addition, the degenerate states do not suffer from any bit flip errors. Further tuning can make the computation topological, thus making the phase resistant to very general errors *(194)*.

**Acknowledgements**

This work was supported by Canada Research Chair (CRC), Canada Foundation for Innovation (CFI), Canada First Excellence Research Fund (CFREF) Program, DFG grants No. MI 658/10-1, No. RO 2247/8-1 and CRC 183, Leverhulme Trust and the Italia-Israel project QUANTRA.

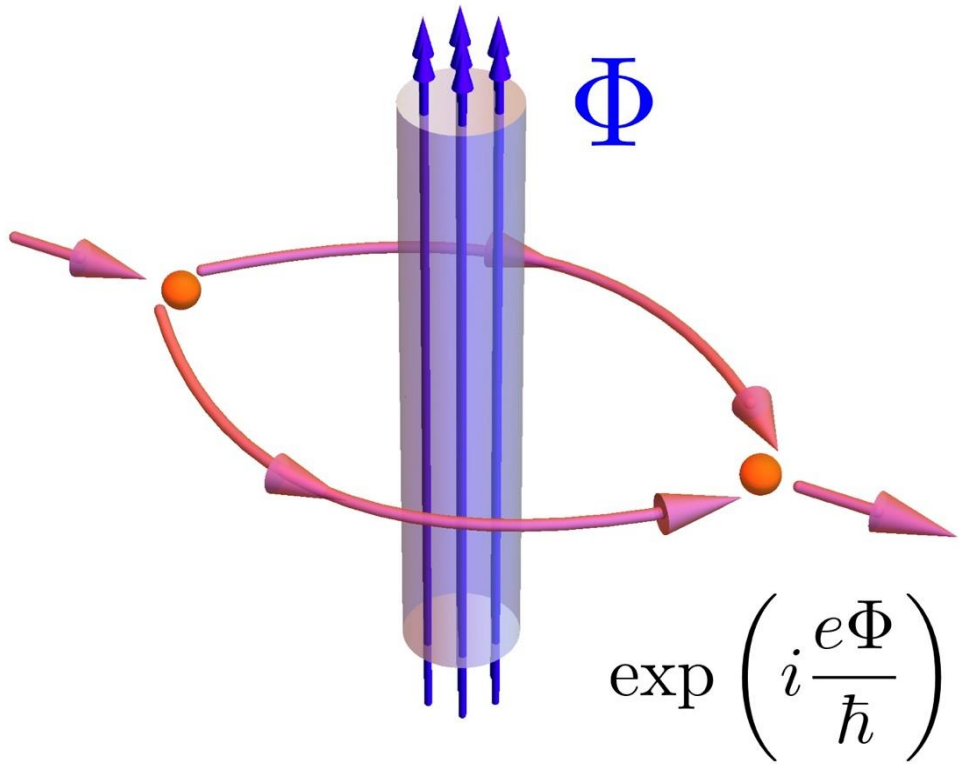

**Fig. 1 | The Aharonov-Bohm effect.** An electron is encircling a magnetic flux Φ (vertical blue arrows) confined to a thin, long solenoid. Although the magnetic field is zero in the vicinity of the superposed wavepackets, the vector potential is non-zero outside the solenoid. Thus, the electronic wavepackets acquire a relative phase of $\exp(ie\Phi/\hbar)$, which causes their interference pattern to change.

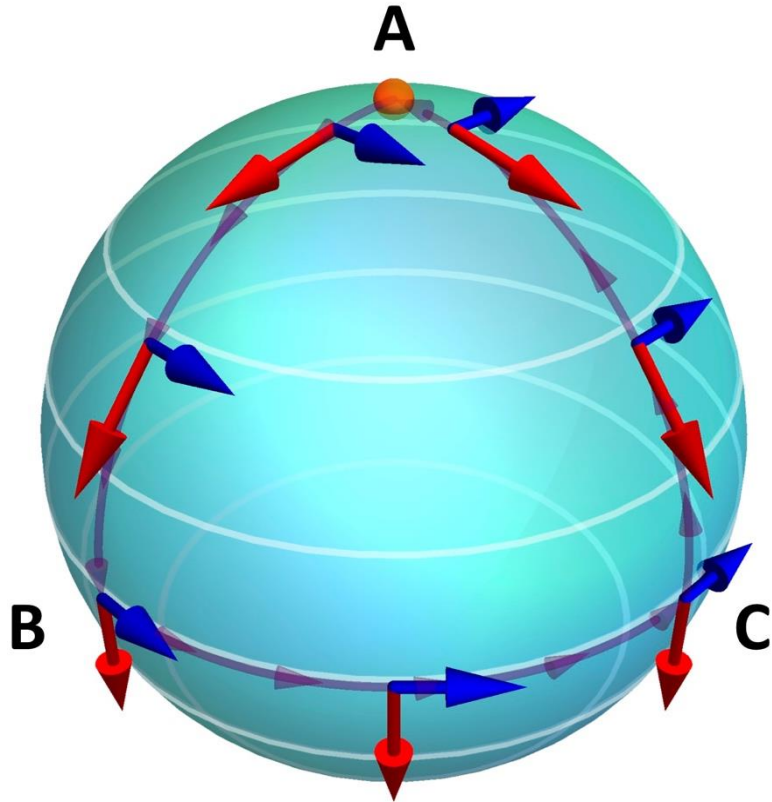

**Fig. 2 | Parallel transport and holonomy within a vector bundle.** The red and blue mutually orthogonal vectors undergo a $\pi/2$- rotation upon $A \to B \to C \to A$. In this depiction, the vector bundle consists of a parameter space $M$ defined by the 2-sphere $S^2$ embedded in $\mathbb{R}^3$ where the wavefunctions are represented by the state vectors, colored in red and blue, that are tangential to the sphere. The connection, $D$, between two points on the sphere effectively changes the Hilbert space that defines the state vector as seen by its rotation upon being parallel transported along the piece-wise smooth path $\gamma$ colored in purple. Upon completing a full-cycle, the state vector experiences a rotation leaving it in a state different from its original one. The rotation experienced by the final state is proportional to half of the enclosed solid angle, where the solid angle here is given by $\pi/2$. This final state can be obtained by applying the holonomy of the path, $H(\gamma,D) = H(\gamma_1,D)H(\gamma_2,D)H(\gamma_3,D)$, to the vector's initial state where $\gamma_1$, $\gamma_2$ and $\gamma_3$ are the three respective segments of the entire path.

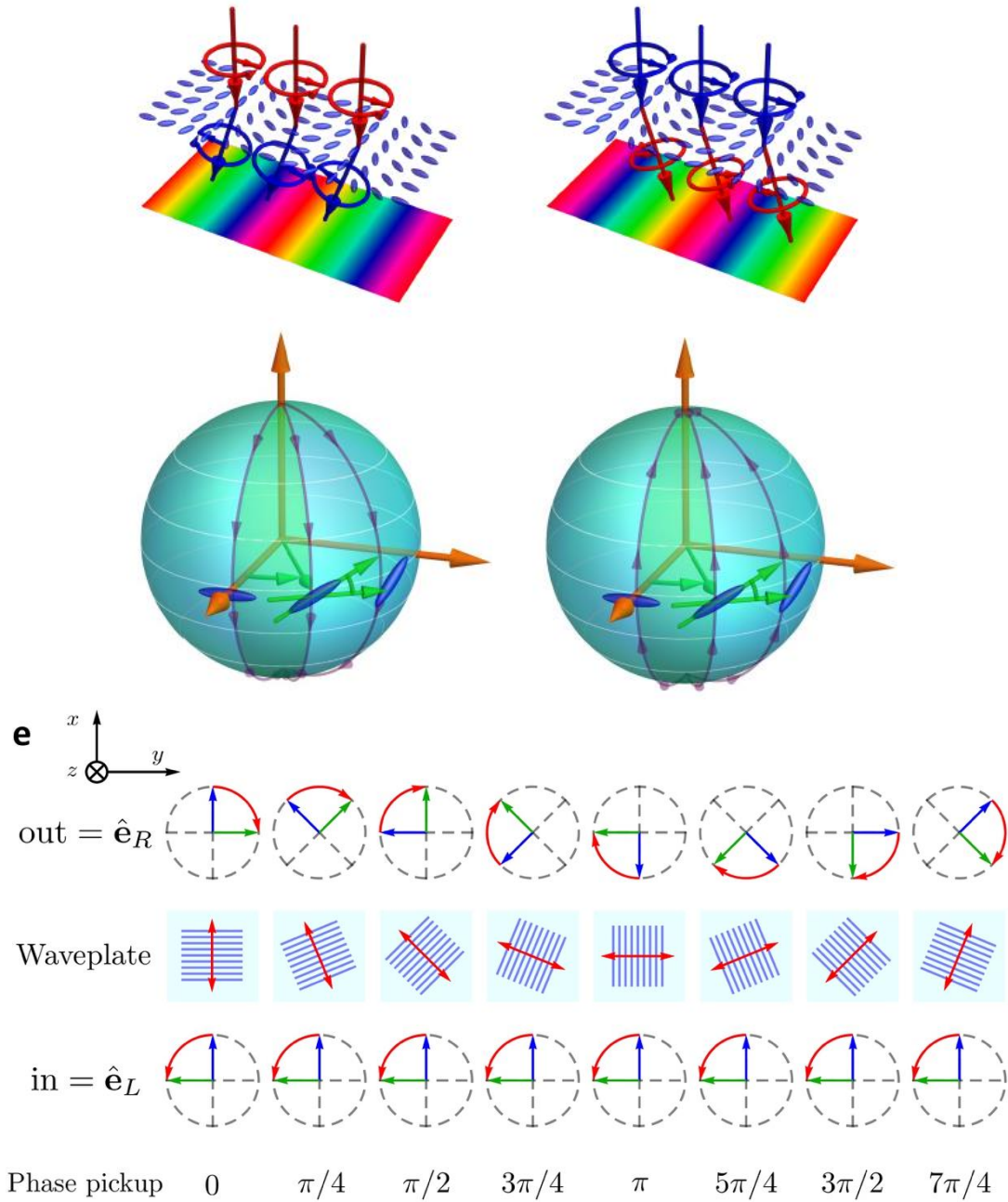

**Fig. 3 | Illustrations of the geometric phase in optics.** In (**a-b**), a circularly polarized light beam passes through a structured birefringent (blue ellipsoids) plate. The induced geometric phase varies in opposite direction for left- $|L\rangle$ and right-polarized $|R\rangle$ light indicated by red and blue arrows, respectively (**a-b**), which causes them to deflect into opposite directions. The incident light gets a space-varying $2\theta$-phase, that is twice the orientation angle of the local optical axis of the structured material, where $\theta$ is coordinate dependent. This geometric phase equals to half of the enclosed solid angle $\Omega$ shown on the polarization Poincaré sphere (**c-d**). The change of sign for left- and right-circular polarizations in the geometric phase is due to the change of orientation in the path taken on the Poincaré sphere. In (**e**), left-polarized light (top row) is passing though half-waveplate elements constructed from nanostructure arrays with their fast axis (straight red arrows) oriented at different angles $\theta$ (middle row). The final state after the half-waveplate will be right-polarized light (bottom row), and acquires a geometric phase that depends on the orientation of the half-waveplates – the corresponding phase pickup for each case is shown in the bottom of the figure. Blue and green arrows indicate electric and magnetic fields, and red arrows show the polarization vectors, respectively.

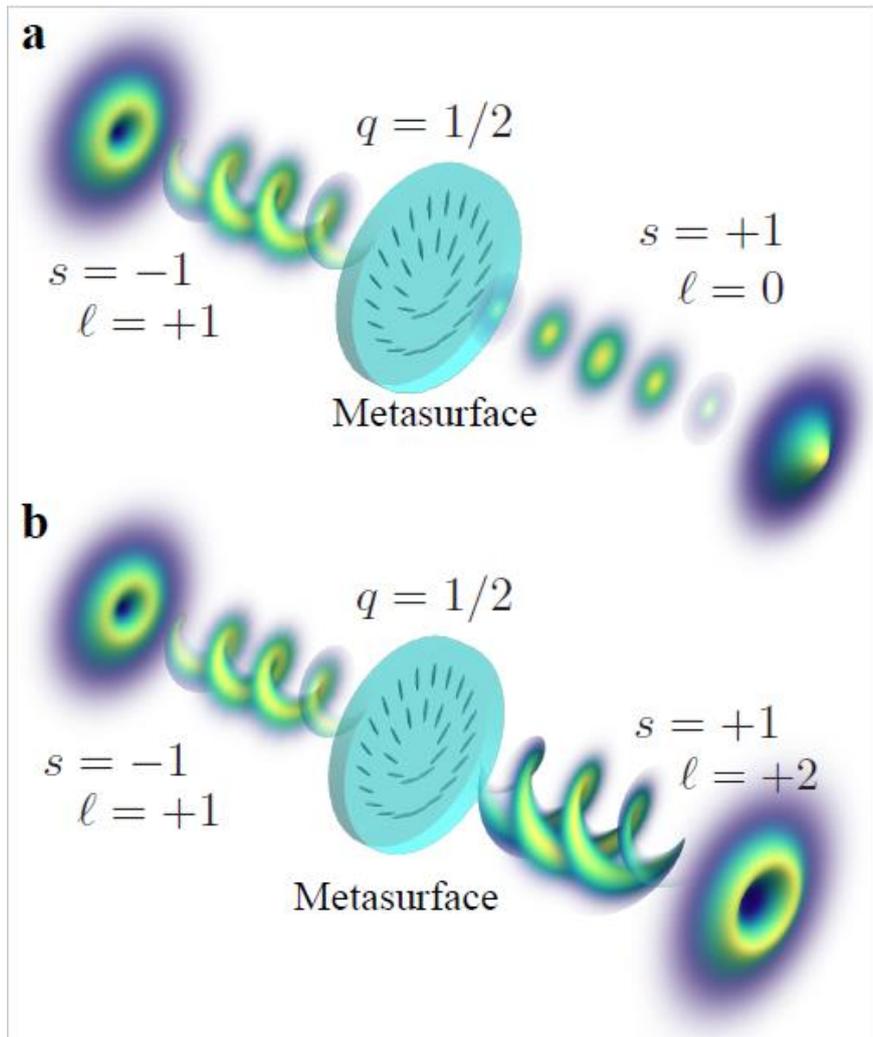

**Fig. 4 | Realizations of a spin-dependent geometric phase through a structured metasurface.** The dielectric metasurface produces a vortex beam with a phase singularity, where $s$ and $\ell$ are the spin and orbital angular momentum, respectively, and $q$ describes the spatial rotation rate of the metasurface structure. Depending on the incident spin angular momentum, the vortex phase eliminates (**a**) or enhances (**b**) the phase singularity, thereby giving rise to a spin-dependent geometric phase. The spin-orbit coupling near the singularity of the geometric phase may lead to the photonic spin Hall effect.

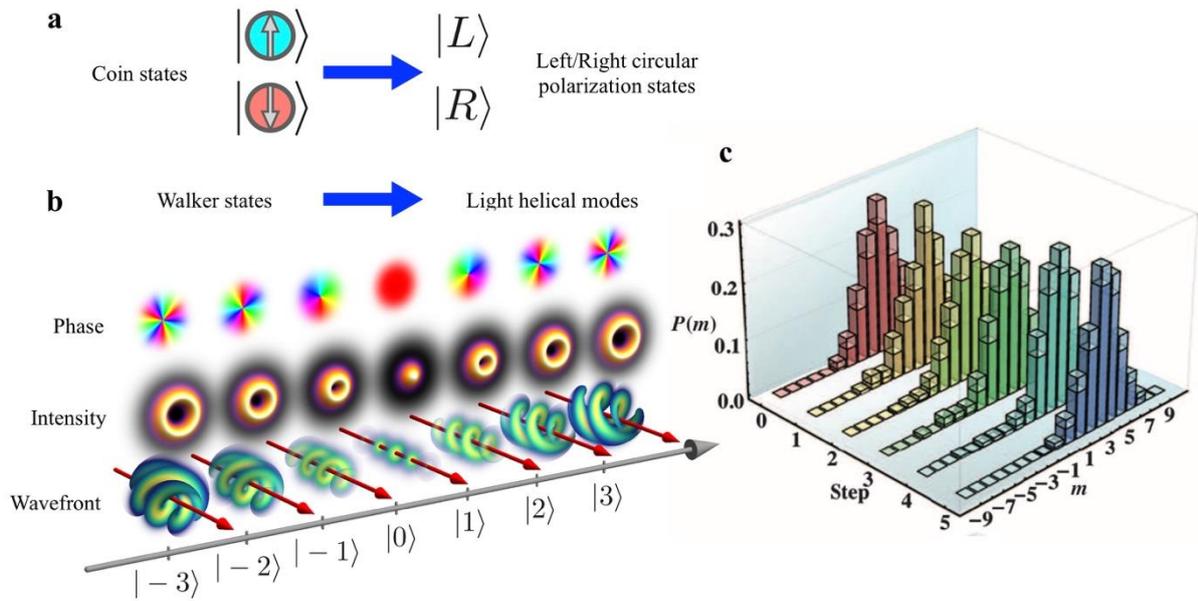

**Fig. 5 | Photonic quantum walk employing orbital angular momentum**. The quantum state describing the polarization of the photons replaces the classical coin (with left and right circular polarization corresponding to 'Heads' and 'Tails') (**a**) and the orbital angular momentum states (spanning from *m=-3* to *m=+3*) replace the classical walker steps (**b**). The phase, intensity and wavefront are sketched for each state. (**c**) An example of experimental simulation of wavepacket dynamics, propagating with a positive group velocity in a five-step quantum walk *(147)*, where different colors are used for denoting different steps.

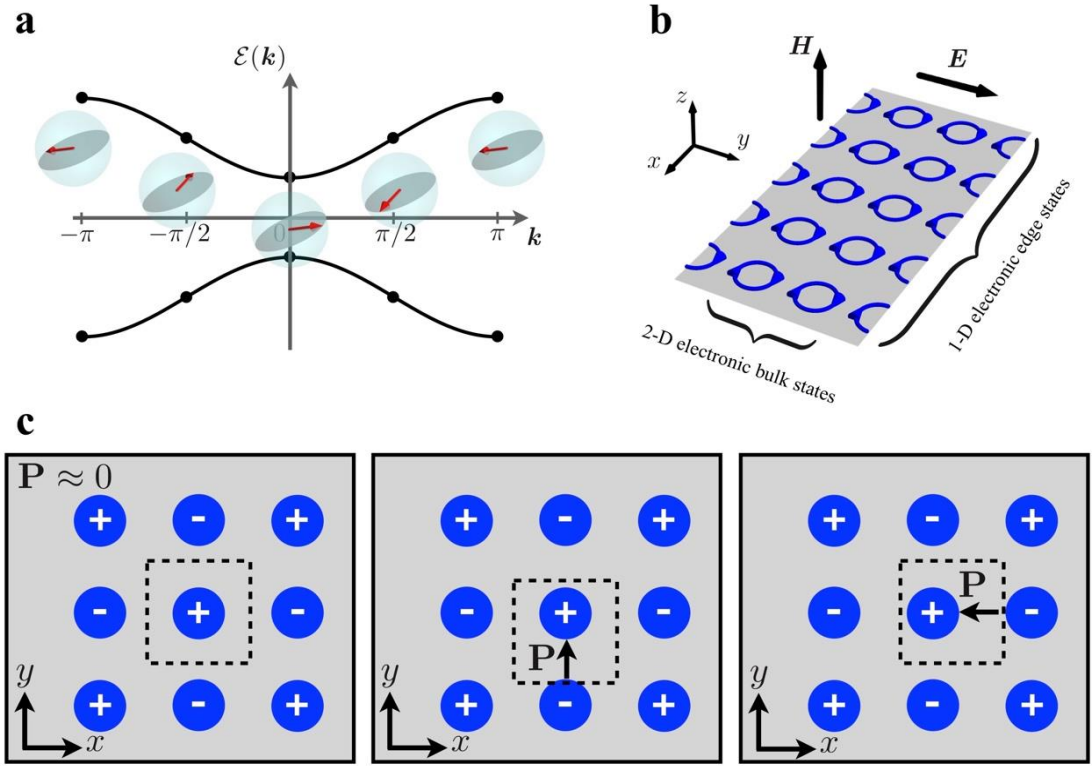

**Fig. 6 | Electronic Bloch states in the 1st Brillouin zone, and the quantum Hall effect.** (**a**) A typical effective band structure of a single electron in a one-dimensional lattice with a lattice constant of unity. As the quasi-momentum $k$ spans the first Brillouin zone , the electron's spin state traverses a (closed) trajectory over the Bloch sphere. The topology of the system may lead to integer or fractional winding numbers which are directly proportional to the resulting Zak phase. (**b**) The quantum Hall effect in a two-dimensional system within the semi-classical picture. The insulating states occur in the bulk and are represented by the closed loop trajectories of the quantized electronic states. The electronic states along the two edges are forced to bounce off the boundaries moving in opposite directions, leading to the edge states. $H$ and $E$ are the external magnetic and electric fields applied to the system, respectively. (**c**) The electric polarization vector in the bulk of a 2D ionic crystal. The magnitude and direction of **P** in the bulk of a 2D crystal is ill-defined and depends on the orientation of the unit cell (shown with dashed lines). Three schematic cases describe (from left to right) a net null polarization, a finite polarization along the vertical axis, and a non-zero polarization vector along the horizontal axis of the crystal.

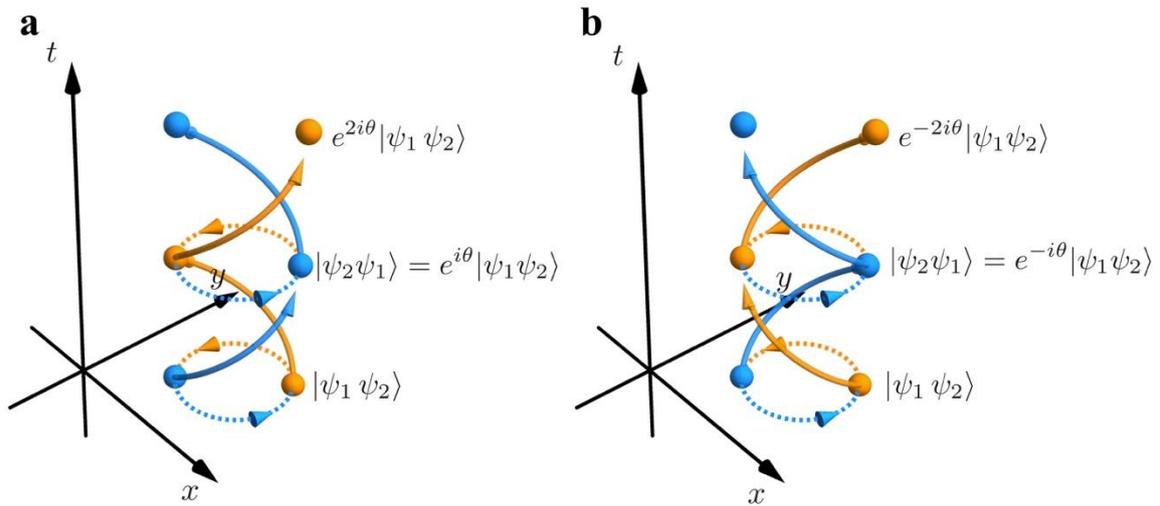

**Fig. 7 | Exchange statistics in the case of anyons.** Two anyons (yellow and blue) switch their places twice in a counter-clockwise manner (**a**) and in a clockwise manner (**b**). The accumulated phase therefore has an opposite sign. The case of bosons/fermions corresponds to $\theta = 0/\theta = \pi$, respectively. In the case of an exchange of anyons, details of the exchange trajectory (that is, the 'history' of the quasi-particles involved) matters. $\psi_1$ and $\psi_2$ are the wavefunctions of the two anyons, $\theta$ is the accumulated phase

| Phase | First appeared in | Mostly known in | Parameter space | Topological | Adiabatic |
|---|---|---|---|---|---|
| Pancharatnam | 1956 | Optics | Poincaré sphere | No | Yes |
| Aharonov-Bohm | 1959 | Quantum electrodynamics | Spacetime | Yes | No |
| Exchange statistics (of Abelian anyons) | 1977 1982 1984 | Condensed matter | Real space | Yes | Yes |
| Berry | 1983 1984 | Quantum mechanics | General | No | Yes |
| Aharonov-Casher | 1984 | Quantum electrodynamics | Real space | Yes | No |
| Hannay angle | 1985 | Classical mechanics | Real space | No | Yes |
| Aharonov-Anandan | 1987 | Quantum mechanics | General | Yes | No |
| Zak | 1989 | Condensed matter | Momentum space | No | No |

**Table 1 | List of some well-known manifestations of the geometric phase.**